\documentclass[epjST]{svjour}
\usepackage{graphics}
\usepackage{epsfig}
\usepackage{color}

\def\beqn{\begin{eqnarray}}
\def\eeqn{\end{eqnarray}}

\begin{document}
\title{Nucleon Polarizabilities: Theory}
\author{B. Pasquini\inst{1}\fnmsep\thanks{\email{pasquini@pv.infn.it}},
D. Drechsel\inst{2}\fnmsep\thanks{\email{drechsel@kph.uni-mainz.de}},
\and
M. Vanderhaeghen \inst{2}
\fnmsep\thanks{\email{marcvdh@kph.uni-mainz.de}}\ }
\institute{ Dipartimento di Fisica Nucleare e Teorica, Universit\`a degli Studi di Pavia, \\and INFN, Sezione di Pavia, I-27100 Pavia,
Italy\and Institut f\"ur Kernphysik, Johannes Gutenberg-Universit\"at, D-55099 Mainz, Germany}
\abstract{ We review recent developments in the theoretical investigation of the nucleon polarizabilities.
We first report on the static polarizabilities as measured
in real Compton scattering, comparing and interpreting the results from various theoretical approaches.
In a second step, we extend the discussion to the generalized polarizabilities
which can be accessed in virtual Compton scattering, showing how the information
encoded in these quantities can provide a spatial interpretation of the
induced polarization densities in the nucleon.}
\maketitle
\section{Introduction}
\label{section:1}

The polarizabilities of a composite system such as the nucleon are
elementary structure constants, just as its size and shape.
They can be accessed experimentally by  Compton scattering processes.
In the case of real Compton scattering (RCS), the incoming real photon deforms
the nucleon, and by measuring the energy and angular distributions of
the outgoing photon one can determine the induced current and magnetization
densities. The global strength of these densities is characterized by
the nucleon dipole and higher order (quasi-static) polarizabilities.  In contrast,
the virtual Compton scattering (VCS) process is obtained if the incident real photon is
replaced by a virtual photon. This
virtuality of the photon gives access to the generalized polarizabilities (GPs) and allows us to map out the spatial
distribution of the polarization densities.
\\
Over the past years, the nucleon polarizabilities have been the subject of extensive experimental and
theoretical studies (see, e.g., Refs.~\cite{Drechsel:2002ar,Drechsel:2007sq,Guichon:1998xv,Schumacher:2005an}
for comprehensive reviews).
In this work we summarize recent developments of the theoretical studies, and we refer to~\cite{fonvieille} for the corresponding analysis
of the experimental work.
In Sec.~\ref{section:2} we discuss the static polarizabilities and compare
the predictions from different theoretical investigations. Several approaches to extract the polarizabilities from  RCS data are
presented in Sec.~\ref{section:3}. The VCS process and the role of the GPs are reviewed in Sec.~\ref{section:4},
and in Sec.~\ref{section:5} we discuss how to obtain a spatial representation of
the information contained in the GPs. A short summary and some conclusions are given in Sec.~\ref{section:6}.
\section{Polarizabilities in RCS}
\label{section:2}
The physical content of the nucleon polarizabilities can be visualized
best by effective multipole interactions for the coupling of the
electric $(\vec{E})$ and magnetic $(\vec{H})$ fields of a photon
with the internal structure of the nucleon.
When expanding the RCS amplitude in the energy of the
photon, the zeroth and first order terms follow from a low-energy
theorem and can be expressed solely in terms of the charge, mass,
and anomalous magnetic moment of the nucleon. The second order terms
in the photon energy describe the response of the nucleon's internal
structure to an electric or magnetic dipole field. They are given by
the following effective interaction~:
\begin{eqnarray}
\label{h2order}
H_{\rm{eff}}^{(2)} = -4\pi \left[
{\textstyle\frac{1}{2}}\,\alpha_{E1}\,\vec{E}\,^2 +
{\textstyle\frac{1}{2}}\,\beta_{M1}\,\vec{H}\,^2\ \right],
\end{eqnarray}
where the proportionality coefficients are the electric
($\alpha_{E1})$ and magnetic ($\beta_{M1}$) scalar dipole polarizabilities,
respectively. These global structure coefficients are proportional
to the electric and magnetic dipole moments of the nucleon which are
induced by the applied electric and magnetic
fields. The polarizabilities have been measured extensively by use of unpolarized
Compton scattering. A global fit to all modern low-energy proton Compton
scattering data leads to the results~\cite{OlmosdeLeon:2001zn}
\begin{eqnarray}
\alpha_{E1}^p & = &  12.1  \pm
0.3 \, (\rm{stat.}) \mp 0.4 \, (\rm{syst.}) \pm 0.3 \, (\rm{mod.}) \ ,
\nonumber \\
\beta_{M1}^p & = & 1.6 \pm 0.4 \, (\rm{stat.}) \pm 0.4 \, (\rm{syst.}) \pm
0.4 \,(\rm{mod.})\ .
\label{eq:exp}
\end{eqnarray}
Here and in the following the scalar polarizabilities are given in units of $10^{-4}$ fm$^3$, and
the indicated errors denote the statistical, systematical and model-dependent errors.
Equation~\ref{eq:exp} shows
the dominance of the electric polarizability $\alpha_{E1}^p$. The tiny value of the
magnetic polarizability, $\beta_{M1}^p$,  originates from a strong
cancelation between the large paramagnetic contribution of the $N \to
\Delta$ spin-flip transition and a nearly equally large
diamagnetic contribution, mostly due to pion-loop effects.
\newline
\indent The internal spin structure of the nucleon appears at third
order in an expansion of the Compton scattering amplitude. It is
described by the effective interaction
\begin{eqnarray}
\label{h3order}
{H}_{\rm{eff}}^{(3)} &=&  - 4 \pi \left[
{\textstyle\frac{1}{2}}\gamma_{E1E1}
\,\vec{\sigma}\cdot(\vec{E}\times\dot{\vec{E}})
+ {\textstyle\frac{1}{2}}\gamma_{M1M1}\,
\vec{\sigma}\cdot(\vec{H}\times\dot{\vec{H}}) \right. \nonumber \\
&&\left. \hspace{.75cm}
- \gamma_{M1E2}\, E_{ij}\,\sigma_iH_j
+ \gamma_{E1M2}\, H_{ij}\,\sigma_iE_j\  \right],
\end{eqnarray}
which involves one derivative of the fields with regard to either
time or space, $\dot{\vec{E}} = \partial_t\vec{E}$ and
$E_{ij}=\frac{1}{2}(\nabla_iE_j+\nabla_jE_i)$, respectively. The
four spin (or vector) polarizabilities $\gamma_{E1E1}$, $\gamma_{M1M1}$,
$\gamma_{M1E2}$, and $\gamma_{E1M2}$ describing the nucleon spin response
at third order, can be related to a multipole
expansion~\cite{Babusci:1998ww}, as is reflected in the subscript notation. For
example, $\gamma_{M1E2}$ corresponds to the excitation of the
nucleon by an electric quadrupole ($E2$) field and its de-excitation by a
magnetic dipole ($M1$) field. Expanding the Compton scattering amplitude to
higher orders in the energy, one obtains higher order
polarizabilities to the respective order, e.g., the quadrupole
polarizabilities at fourth order~\cite{Babusci:1998ww,Holstein:1999uu}.

On the experimental side, much less is known about the spin polarizabilities,
except for the forward ($\gamma_0$) and backward
($\gamma_{\pi}$) spin polarizabilities of the proton, given
by the following linear combinations of the polarizabilities in
Eq.~(\ref{h3order}):
\begin{eqnarray}
\gamma_{0}&=&-\gamma_{E1E1}-\gamma_{M1M1}-\gamma_{E1M2}-\gamma_{M1E2}
\,,\label{eq:g0_mult}\\
\gamma_{\pi}&=&-\gamma_{E1E1}+\gamma_{M1M1}-\gamma_{E1M2}+\gamma_{M1E2}
\,.\label{eq:gpi_mult}
\end{eqnarray}
The forward spin polarizability has been determined by the
Gerasimov-Drell-Hearn sum rule experiments at MAMI and
ELSA~\cite{Ahrens:2001qt,Dutz:2003mm,Dutz:2004}. A recent analysis of these data
yields the value~\cite{Pasquini:2010zr}
\begin{eqnarray}
\label{eq:go} \gamma_0=-0.90\pm 0.08  \, (\rm{stat.})\pm 0.11 \, (\rm{syst.})\, ,
\end{eqnarray}
where here and in the following all spin polarizabilities are given
in units of $10^{-4}$ fm$^4$.
The recent experimental value for the backward spin
polarizability has been obtained by a dispersive analysis of
backward-angle Compton scattering~\cite{Schumacher:2005an},
\begin{eqnarray}
\gamma_\pi=-38.7 \pm
1.8 \, (\rm{stat.+syst.}) \, .\label{eq:gpi}
\end{eqnarray}
This value includes both the dispersive and the large $\pi^0$-pole contributions.
In the analysis of Ref.~\cite{Schumacher:2005an}, the latter contribution takes the value
$\gamma_\pi^{\pi^0-{\rm pole}}=-46.7$, which leads to $\gamma_\pi^{{\rm disp}}=8.0\pm 1.8$.

The effective Hamiltonians of Eqs.~(\ref{h2order}) and
(\ref{h3order}) describe a shift in the nucleon energies at second
order in the electromagnetic (e.m.) fields. Several lattice QCD collaborations have recently
implemented this fact as a tool to   ``measure'' the nucleon polarizabilities by calculating
the mass shifts in a constant background field and isolating the quadratic response. In particular, it has been
possible to determine the electric polarizability of the neutron and
other neutral octet and decuplet baryons~\cite{Christensen:2004ca} as well as
the magnetic polarizability of the proton, the neutron, and all
other particles in the lowest baryon octet and decuplet
states~\cite{Lee:2005dq}. In Ref.~\cite{Lee:2005dq},
$\beta_{M1}^p$ has been calculated by use of the
Wilson action in the pion mass range
$0.5 \leq m_\pi \leq 1$~GeV and neglecting disconnected loop diagrams.
For the smallest calculated pion mass
of $m_\pi \simeq 500$~MeV, a value of $\beta_{M1}^p = 2.36 \pm
1.20 $ was obtained. While it is
encouraging to see that the lattice result is in the right ball park
when compared to the experimental value of Eq.~(\ref{eq:exp}), a
more precise comparison clearly requires a dynamical fermion
calculation, including disconnected loop diagrams and much smaller
pion masses. Such small pion masses would then allow one to
extrapolate safely to the physical pion mass within the framework of
chiral perturbation theory.

To determine the spin polarizabilities, Eq.~(\ref{h3order})
can likewise be used to calculate energy shifts of a polarized
nucleon in an external field. As an example, consider a nucleon
polarized along the $z$-axis and apply a magnetic field rotating
with angular frequency $\omega$ in the $xy$ plane,
\begin{eqnarray}
\vec H = B_0 \left[ \cos (\omega t) \; \hat e_x + \sin (\omega t) \; \hat e_y
\right],
\end{eqnarray}
where $\hat e_i$ stands for the unit vector in the direction $i = x,
y$ and $B_0$ is the magnitude of the field. Such a field leads to
an energy shift $\Delta E = \mp 2 \pi \gamma_{M1 M1} \omega B_0^2$
if the nucleon spin is oriented along the positive ($-$) or negative
($+$) $z$-axis. The split between the two levels is then directly
proportional to the magnetic dipole spin polarizability
$\gamma_{M1M1}$. It has been shown in
Ref.~\cite{Detmold:2006vu} that allowing for
background fields with suitable variations in space and time,
lattice QCD should be able to calculate all six dipole
polarizabilities. In particular, calculations are in progress to
determine the electric polarizability of a charged particle such as
the proton as well as the four proton spin polarizabilities of
Eq.~(\ref{h3order})~\cite{Detmold:2006xh}.
\newline
\indent
A microscopic understanding of the nucleon's
polarizabilities requires to quantify the interplay between
resonance contributions, e.g., the $N \to \Delta$ transition, and
long range pion-cloud effects. Systematic
studies of such pion-cloud effects became possible with the development of chiral
perturbation theory (ChPT), by an expansion of the
lagrangian in the external momenta and the pion mass
(``$p$-expansion''). The first such calculation at leading order, $\mathcal{O}(p^3)$,
yielded the following values for the
proton scalar polarizabilities~\cite{Bernard:1991rq}:
\begin{eqnarray}
\alpha_{E1}^p = 10\beta_{M1}^p = \frac
{5\alpha_{em}g_A^2}{96\pi f_{\pi}^2m_{\pi}} = 12.2\ ,
\label{eq:lo}
\end{eqnarray}
where $\alpha_{em} = 1/137$, $g_A \simeq 1.27$, and $f_\pi =
92.4$~MeV. This result is in remarkable agreement with the
experimental result of Eq.~(\ref{eq:exp}). It also illustrates that
these quantities diverge in the chiral limit, which is a challenge
for the lattice QCD calculations. Conversely, the chiral expansion converges well in the
``small $m_\pi$'' regime, and therefore
ChPT can complement the lattice calculations by extrapolation to the
physical pion mass. The ChPT work was extended to $\mathcal{O}(p^4)$
within the heavy-baryon expansion~\cite{Bernard:1993bg}, yielding
$\alpha_{E1}^p=10.5 \pm 2.0$ and $\beta_{M1}^p=3.5\pm 3.6$.
The error bars for these values indicate that several low-energy constants had to be determined
by resonance saturation, e.g., by putting in phenomenological information about resonance and vector mesons.
Since the $\Delta(1232)$ lies close in energy, it has been proposed to include the resonance dynamically.
This leads to an additional expansion parameter, the $N\Delta$ mass splitting ($\epsilon$~expansion).
Unfortunately, at ${\cal O}(\varepsilon^3)$ the ``dynamical'' $\Delta$ increases the polarizabilities
to values far above the data, $\alpha_{E1}^p=16.4$ and $\beta_{M1}^p=9.1$. This can be changed
by introducing large low-energy constants within a higher-order calculation, however, at expense
of losing the predictive power~\cite{Hildebrandt:2003fm}.
More recently Pascalutsa and Phillips~\cite{Pascalutsa:2002pi} proposed an alternative treatment
in which the power counting changes between the threshold and the resonance regions, depending on which scales are enhanced.
At low energies they introduce the ratio $\delta$, with 
$\delta=(M_\Delta-M_N)/M_N$ as new expansion parameter.
In this scheme, the values for the scalar polarizabilities at N$^3$LO are
$\alpha_{E1}^p=10.8\pm 0.7$ and $\beta_{M1}^p=4.0\pm 0.7$, where the error bar is
an estimate of higher-order contributions~\cite{Lensky:2009uv}, and show a much improved agreement with experiment.

\begin{table}
\caption{
Theoretical predictions for the dispersive contribution to spin polarizabilities
of the proton: to ${\cal O}(p^3)$ in HBChPT~\cite{Hemmert:1997tj}, to ${\cal O}(p^4)$
in HBChPT~\cite{VijayaKumar:2000pv},  to ${\cal O}(\varepsilon^3)$ in the small scale expansion~\cite{Gellas:2001yv}, in the Lorentz covariant (LC) ChPT to ${\cal O}(p^4)$~\cite{dalibor},
in the chiral approach with unitarity and causality (GLP) of Ref.~\cite{Gasparyan:2011yw},
in fixed-$t$ dispersion relations analysis (HDPV) of Ref.~\cite{Holstein:1999uu} and
Ref.~\cite{Babusci:1998ww} (BGLMN), and in hyperbolic DRs of Ref.~\cite{Drechsel:2002ar} (HYP. DRs).}
\label{tab:1}       
\begin{tabular}{ccccccccc}
\hline\noalign{\smallskip}
&${\cal O}(p^3)$ & ${\cal O}(p^4)$ &   ${\cal O}(\varepsilon^3)$&
LC & GLP & HDPV & BGLMN & HYP. DRs \\
\noalign{\smallskip}\hline\noalign{\smallskip}
$\gamma_{E1E1}$ & -5.7$\, $ & -1.4 & -5.4  & -2.8      &-3.7  & -4.3 & -3.4 & -3.8\\
$\gamma_{M1M1}$ & -1.1 &  $\, \,3.3$ &  $\, \,1.4 $ & -3.1      &$\, \,2.5 $ & $\, \,2.9 $  &$\, \,2.7$ & $\, \,2.9$\\
$\gamma_{E1M2}$ & $\, \, 1.1$  &  $\, \,0.2 $&  $\, \,1.0$  &  $\, \,0.8$    &$\, \,1.2 $ & -0.01  & $\, \,0.3$ & $\, \,0.5$\\
$\gamma_{M1E2}$ &  $\, \,1.1$  &  $\, \,1.8 $&  $\, \,1.0$  &  $\, \,0.3$  &$\, \,1.2 $ &  $\, \,2.1 $   & $\, \,1.9$ & $\, \,1.6$ \\
$\gamma_0$     &  $\, \,4.6$  & -3.9 &  $\, \,2.0$  &  $\, \,4.8$  &-1.2 & -0.7    &-1.5 & -1.1\\
$\gamma_\pi$   &  $ \,\, 4.6 $ &  $\, \,6.3$ &  $\, \, 6.8$  & -0.8  &$\, \,6.1$ & $\, \,9.3$     &$\, \,7.8 $& $\, \,7.8$\\
\noalign{\smallskip}\hline
\end{tabular}
\end{table}

Results within different theoretical approaches for the dispersive contribution to
the spin polarizabilities are collected in Table~\ref{tab:1}.
The predictions from unsubtracted fixed-$t$ dispersion relations (DRs) are based on different inputs
for the pion photoproduction amplitudes, while the results from hyperbolic DRs
are obtained from dispersion integrals which run along a path at fixed angle.
The agreement between the different DR results in Table~\ref{tab:1} is quite satisfactory in all cases,
and the spread among the predicted values can be seen as the best possible
error estimate of such calculations to date.
The ChPT predictions in the heavy baryon, small-scale expansion and Lorentz-covariant
approach disagree in some cases, both among each other and with the DR results.
In particular, issues about the convergence of the chiral expansion
for the spin polarizabilities deserve further studies.
In Table~\ref{tab:1} we also quote the results from a recent approach based  on  the chiral Lagrangian
at order ${\cal O}(p^3)$ involving pion, nucleon and photon fields, and without
isobar fields as explicit degrees of freedom (see column labeled ``GLP'')~\cite{Gasparyan:2011yw}.
The physics of the isobar resonances is included by infinite summation of higher order counter terms in the chiral Lagrangian,
taking into account the constraints from unitarity and causality.
In this framework, the authors of Ref.~\cite{Gasparyan:2010xz} were able to give an unified description of photon- and
pion-nucleon scattering data up to and beyond the isobar region, with the same set of parameters
adjusted to pion-nucleon scattering data.
For $\gamma_{E1M2}$ and $\gamma_{M1E2}$, the ChPT results at ${\cal O}(p^3)$
are recovered within small numerical deviations (compare columns ${\cal O}(p^3)$ and GLP in Table~\ref{tab:1}),
which are due to different values used for the parameters used in the calculations.
In contrast, the ChPT and GLP results for $\gamma_{E1E1}$ and $\gamma_{M1M1}$  are at variance because of
large rescattering effects, and the GLP values are in much better agreement with the predictions from DRs.
The GLP scheme shows distinct indications of an improved convergence pattern as
compared to ChPT, as was also observed for neutral-pion photoproduction~\cite{Gasparyan:2010xz}.
However, these expectations for the spin polarizabilities need to be confirmed by explicit calculations.
\section{Extraction of RCS polarizabilities}
\label{section:3}
The extraction of polarizabilities from RCS
data has been performed by three techniques. The first one is a low-energy expansion (LEX) of the Compton cross sections. Unfortunately this
procedure is only applicable at photon energies well below 100~MeV,
which makes a precise extraction a rather challenging task. The
sensitivity to the polarizabilities is increased by measuring
Compton scattering observables around pion threshold and into the
$\Delta(1232)$ resonance region.
A second formalism which has been successfully applied to Compton data
is based on DRs, in particular unsubtracted~\cite{L'vov:1996xd} and
subtracted~\cite{Drechsel:1999rf} fixed-$t$ DRs were used to extract the scalar polarizabilities
of Eq.~(\ref{eq:exp}).
A third approach has been developed within the framework of chiral effective field
theories~\cite{Hildebrandt:2003fm,Pascalutsa:2002pi,Lensky:2009uv,Beane:2004ra,Hildebrandt:2003md}.
For energies below and around pion threshold the full Compton scattering process has been calculated
to $\mathcal{O}(p^4)$, allowing for an independent extraction
of the polarizabilities from Compton scattering data. The thus
obtained values for $\alpha_{E1}^p$ and $\beta_{M1}^p$
in the work of \cite{Hildebrandt:2003fm,Beane:2004ra} are nicely
compatible with the results given by Eq.~(\ref{eq:exp}).
Recently, the ``$\delta$-counting''  scheme has been applied to extract the scalar polarizabilities using
a subset of low-energy data up to $E_\gamma=149 $ MeV~\cite{Lensky:2009uv}, and a preliminary
study in a larger energy range has been presented in~\cite{McGovern:2009sw}.
Both these analyses tend to favor a larger value for $\beta_{M1}^p$
than given by Eq.~(\ref{eq:exp}). These findings call for new dedicated  measurements below pion
threshold. In particular, it
has been suggested that the sensitivity to the scalar polarizabilities can be enhanced by
combination of unpolarized cross sections and data obtained
with linearly polarized photons in particular kinematics.
Proposals for such measurements below pion threshold have been presented at the HI$\gamma$S facility at Duke~\cite{Weller:2009zza}.

Because a reliable data analysis is based on DRs, we
recall some pertinent features of this technique in the following. The
$T$ matrix of RCS can be expressed by 6 independent structure functions $A_i(\nu,t)$~\cite{L'vov:1996xd}.
These functions depend on the variables $\nu$ and $t$, which are related to the
initial ($E_\gamma$) and final ($E'_\gamma$) photon laboratory energies, and
to the scattering angle $\theta_{\rm{lab}}$ by $t=-4E_\gamma E'_\gamma\sin^2(\theta_{\rm{lab}}/2) $
and $\nu=E_\gamma+t/(4M_N)$.
The invariant amplitudes $A_i$ are free of kinematical singularities and
constraints, they also obey crossing symmetry and gauge invariance.
Assuming further analyticity and an appropriate high-energy behavior, these
amplitudes fulfill unsubtracted DRs at fixed $t$,
\begin{equation}\label{eq:4.1.17}
{\rm {Re}}~A_i(\nu, t) = A_i^{\rm{pole}}(\nu, t) +\frac {2}{\pi} \, \mathcal
{P} \int_{\nu_{0}}^{\infty} d\nu' \; \frac{\nu' \, {\rm {Im}}~A_i(\nu',t)} {
\nu'^2 - \nu^2}\,,
\end{equation}
where $A_i^{\rm{pole}}$ is the nucleon pole term, $\nu_0$ is the pion production threshold and $\mathcal {P}$ denotes the
principal value integral. The latter can be calculated if the absorptive part
of the amplitude, Im~$A_i$, is known to a sufficient accuracy. Because of the
energy weighting, the pion production near threshold and the mesonic decay of
the low-lying resonances yield the biggest contributions to the integral.  With
the existing information on these processes and some reasonable assumptions on
the lesser known higher part of the spectrum, the integrand can be constructed
up to cm energies $W \approx 2$~GeV. However, Regge theory predicts that the
amplitudes $A_1$ and $A_2$ do not drop sufficiently fast to warrant a
convergence of the integral. This behavior is mainly due to fixed poles in the
$t$ channel. In particular the $t$-channel exchange of pions and $\sigma$
mesons leads to the bad convergence for $A_2$ and $A_1$, respectively. The
latter meson has a mass of about 600~MeV and a very large width, it models
correlations in the two-pion channel with spin and isospin zero and positive parity. In
order to obtain useful results for these two amplitudes,
\cite{L'vov:1996xd} proposed to close the contour integral in the complex
plane by a semi-circle of finite radius $\nu_{max}$, and to replace the
contribution from the semi-circle by a number of energy independent poles in
the $t$ channel. This procedure is relatively safe for $A_2$ because the
$\pi^0$ pole or triangle anomaly is well established by both experiment and
theory. However, it introduces a considerable model-dependence for $A_1$.
\newline
\indent
In order to avoid the convergence problem and the phenomenology necessary to
determine the asymptotic contributions, it was suggested to subtract the DRs at
$\nu=0$ \cite{Drechsel:1999rf}. This subtraction improves the convergence by
two additional powers of $\nu'$ in the denominator of the dispersion integrals,
Eq.~(\ref{eq:4.1.17}). The subtraction functions $A_i(\nu=0, t)$ can be
obtained from subtracted DRs in $t$ with the imaginary part of the amplitude
$\gamma\gamma\rightarrow\pi\pi\rightarrow N\bar{N}$ as input. In a first step,
a unitarized amplitude for the $\gamma\gamma\rightarrow\pi\pi$ subprocess is
constructed from the available experimental data. This information is then
combined with the $\pi\pi\rightarrow N\bar{N}$ amplitudes determined by
analytic continuation of $\pi N$ scattering amplitudes \cite{Hoehler:1983dd}.
Once the $t$ dependence of the subtraction functions $A_i(0,t)$ is known, the
subtraction constants $a_i=A_i(0, 0)$ have to be fixed.
The 6 subtraction constants $a_1$ to $ a_6$
are given by linear combinations of the scalar and spin polarizabilities
and can be used to fit the RCS data.
In the analysis of unpolarized cross sections it is sufficient
to fit $a_1$ and $a_2$, or equivalently ($\alpha_{E1} -
\beta_{M1}$) and $\gamma_\pi$ to the data. The remaining 4 subtraction
constants can be calculated through an unsubtracted dispersion integral.
\begin{figure}
\begin{center}
\includegraphics[width = 12 cm]{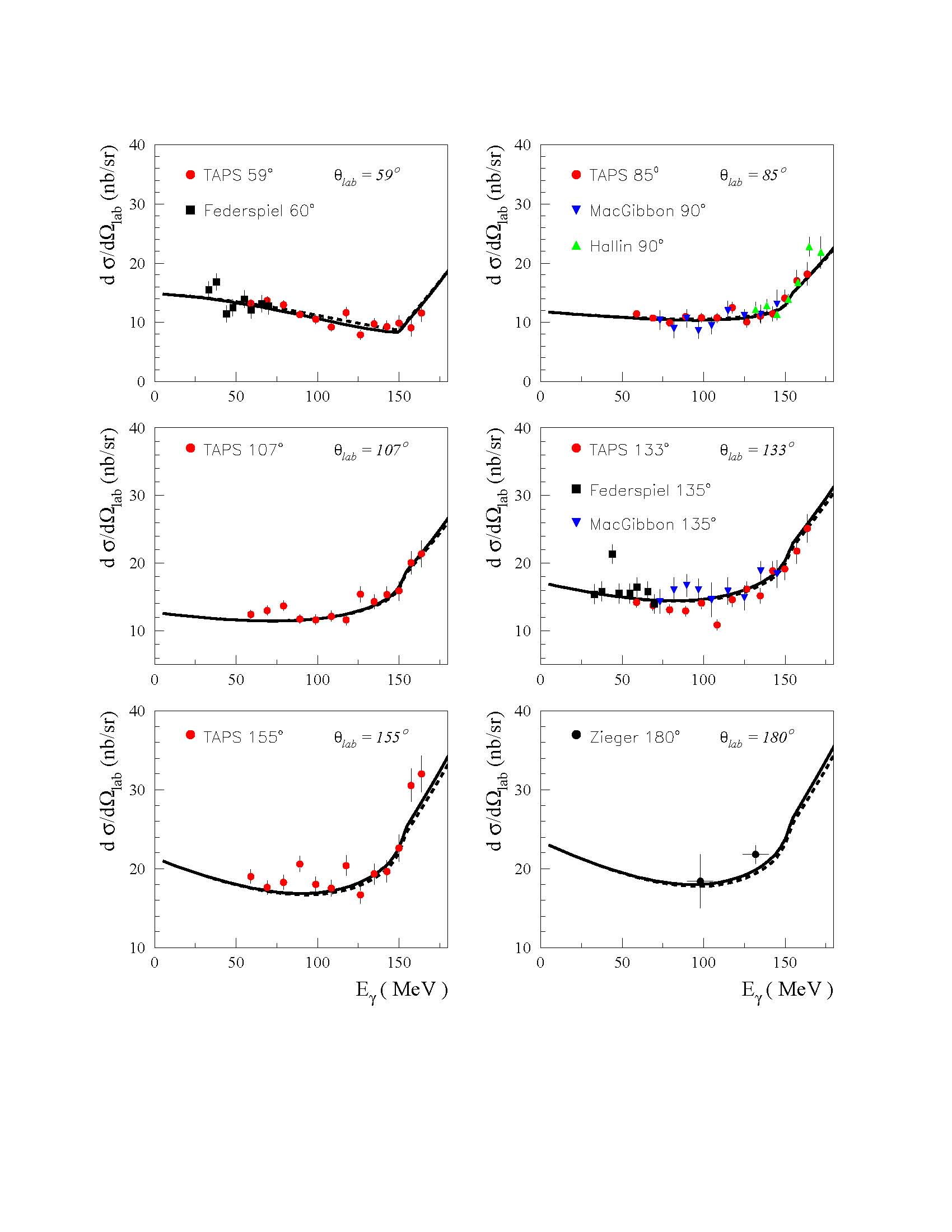}
\end{center}
\vspace{-2.8 cm}
\caption{Differential cross section for Compton scattering off the proton as a function of the
lab photon energy $E_\gamma$ and at different scattering angles.
Solid lines: results from fixed-$t$ subtracted DRs, dashed lines: fixed-$t$ unsubtracted DRs.
All results are shown for fixed values of $\alpha_{E1}+\beta_{M1}=13.8$, $\alpha_{E1}-\beta_{M1}=10$,
and $\gamma_\pi=-37$. The  data are from Ref.~\cite{OlmosdeLeon:2001zn} (full circles),
Ref.~\cite{Federspiel:1991yd} (diamonds), Ref.~\cite{MacGibbon:1995in} (triangles),
and Ref.~\cite{Zieger:1992jq} (open circles).\label{fig1}}
\end{figure}
\newline
\indent
In Fig.~\ref{fig1} we show the differential cross sections measured at several 
laboratories, as a function of
the lab photon energy and at different scattering angles.
The solid and dashed lines correspond to the results from unsubtracted and subtracted fixed-$t$ DRs,
respectively, as obtained for fixed values of $\alpha_{E1}$, $\beta_{M1}$ and $\gamma_\pi$.
Except for extreme backward scattering, both approaches lead to nearly identical results.
A fit to the data within the unsubtracted DR formalism yields the values in Eq.~(\ref{eq:exp}) for the scalar polarizabilities, and
compatible results, within error bars, are also obtained from subtracted DRs~\cite{Drechsel:2002ar}.
Because both formalisms provide an independent cross-check for the extraction of the polarizabilities, we can conclude
that the present analysis of the low-energy data is well under control.
We also note that very similar results for the cross sections below pion threshold were recently obtained
within the unitary approach of Ref.~\cite{Gasparyan:2011yw}.
%
\begin{figure}[t]
\begin{center}
\includegraphics[width = 0.325\columnwidth]{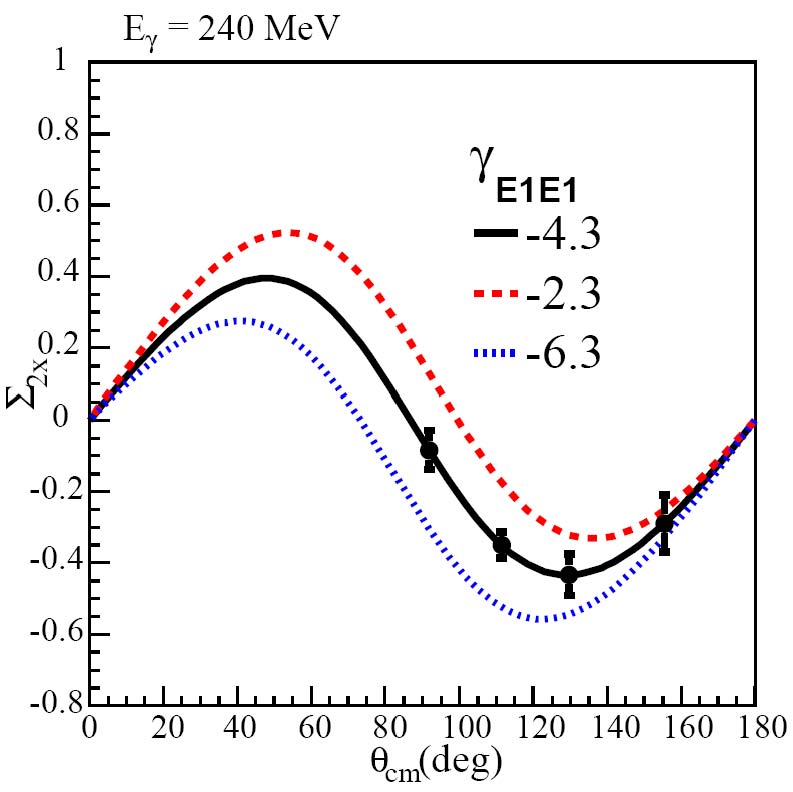}
\includegraphics[width = 0.325\columnwidth]{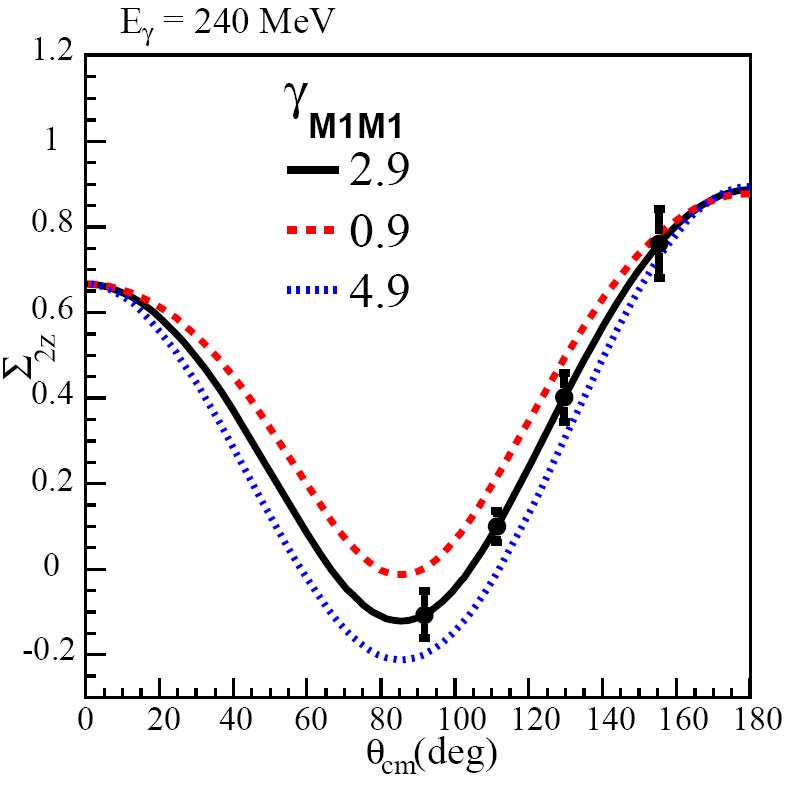}
\vspace{0.1 cm}
\includegraphics[width = 0.325\columnwidth]{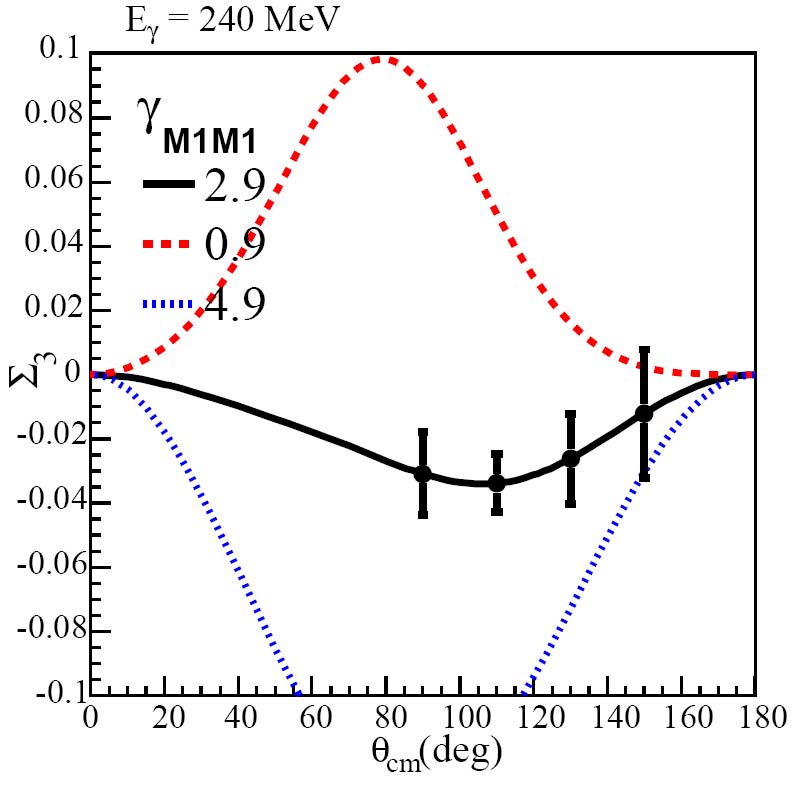}
\end{center}
\caption{The beam-target asymmetries $\Sigma_{2x}$ (left), $\Sigma_{2z}$ 
(center),
and the beam asymmetry $\Sigma_3$ (right), as function of the cm scattering angle at fixed photon energy $E_\gamma=240$ MeV.
The different lines show the predictions of subtracted DRs using the experimental values for
$\alpha_{E1}$, $\beta_{M1}$, $\gamma_0$, and $\gamma_\pi$ as given by Eqs.~(\ref{eq:exp}), (\ref{eq:go}) and (\ref{eq:gpi}),
while the remaining polarizabilities are free parameters.
Left panel: results for fixed $\gamma_{M1M1}=2.9$  and values of $\gamma_{E1E1}$ as indicated.
Central and right panels: results for fixed $\gamma_{E1E1}=-4.3$ and values of $\gamma_{M1M1}$ as indicated.
The plots are taken from the MAMI proposal~\cite{MAMI:2009dd}, with the points representing simulated data
with the expected error bars for 100 hours running time.\label{fig2}}
\end{figure}
\newline
\indent
Because the unpolarized cross section is mainly sensitive to the scalar polarizabilities and to the
backward spin polarizability, double polarization experiments  will be required
to get information
on the individual spin polarizabilities.
First measurements of these observables  have recently started at MAMI~\cite{MAMI:2009dd}.
The proposal is to measure three different observables, $i)$ the photon-beam asymmetry $\Sigma_3$
with linearly polarized photons, $ii)$ the beam-target asymmetry $\Sigma_{2z}$ with circularly polarized photons
and a longitudinally polarized target, and $iii)$ the beam-target asymmetry $\Sigma_{2x}$ with
circularly polarized photons and a transversely polarized target.
The simulated data points with the expected error bars for 100 hours running time are shown in Fig~\ref{fig2},
in comparison with the predictions from subtracted DRs obtained for different values of  $\gamma_{E1E1}$  and $\gamma_{M1M1}$~\cite{Pasquini:2007hf}.
From a preliminary analysis in Ref.~\cite{MAMI:2009dd}, the  projected errors of the extracted spin polarizabilities
are expected to be smaller than the spread of values presented in Table~\ref{tab:1}. Therefore, these experiments
hold the promise to discriminate between the different theoretical predictions.

\begin{figure}[t]
\begin{center}
\epsfig{file=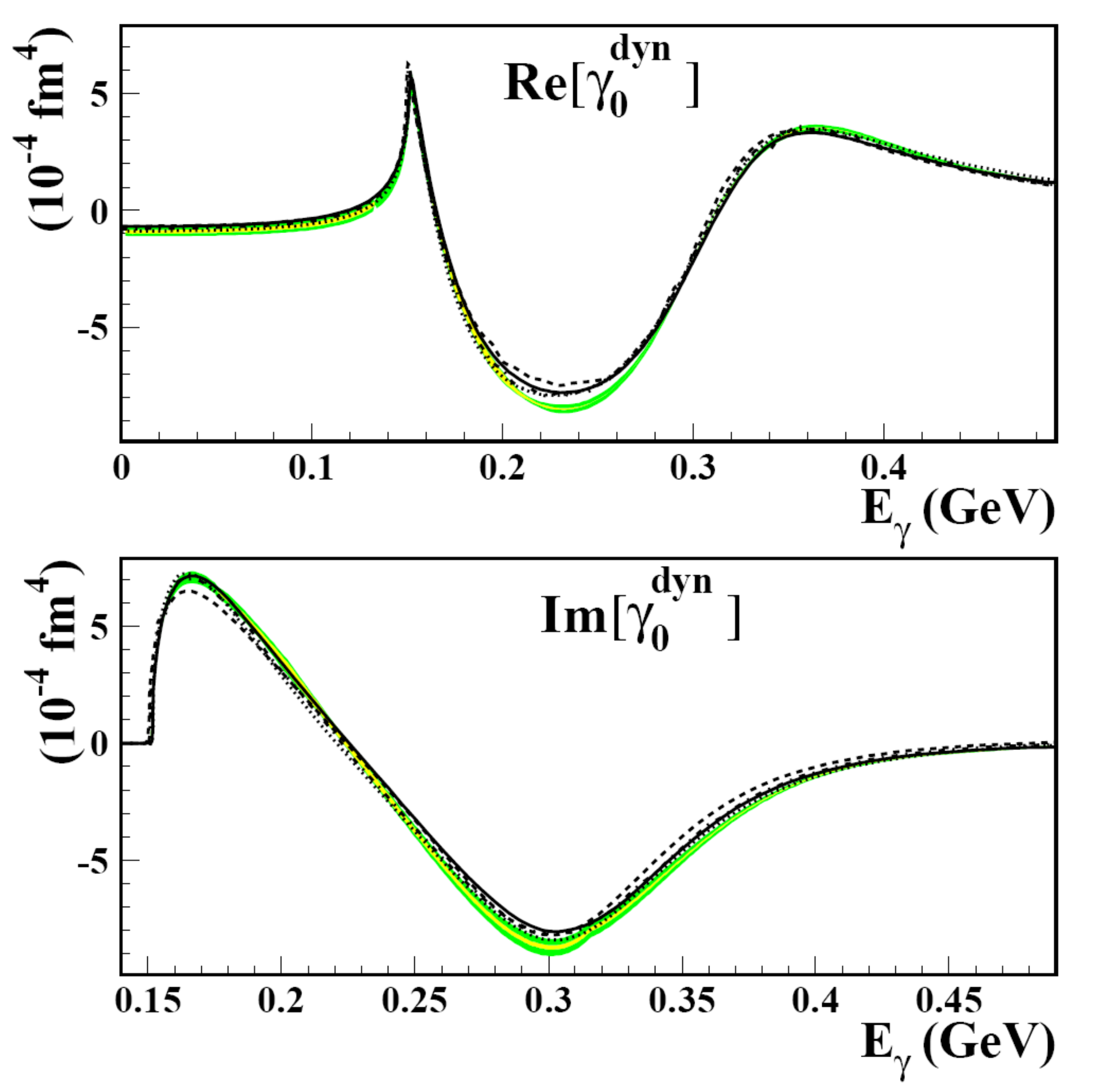,  width=0.60\columnwidth}
\end{center}
\caption{
The real (upper panel) and imaginary (lower panel) parts of the dynamic forward spin polarizability
$\gamma_0^{\rm {dyn}}$ as function of the photon lab energy $E_{\gamma}$.
The light grey (yellow) and dark grey (green) bands show the statistical  and systematic uncertainties.
The predictions are based on the pion photoproduction multipoles of
Hanstein {\emph {et al.}}~\cite{Hanstein:1997tp} (solid line), SAID~\cite{Arndt:2002xv} (dotted line),
 MAID07~\cite{Drechsel:2007if} (dashed line), and DMT~\cite{DMT:2001} (dash-dotted line).\label{fig3}}
\end{figure}

However, all the mentioned approaches for the extraction of the polarizabilities
will necessarily contain some model dependence through parameters describing the high-energy regime. Such parameters are the low-energy
constants appearing in ChPT, and in  DRs the extrapolation of the photoproduction
data to regions not covered by the experiment. Any additional confirmation of the theoretical framework by new alternative theoretical
developments or independent experimental information is therefore most welcome.
\newline
\noindent
To this aim, the DR calculation for the double polarization observables to be measured at MAMI,
was recently compared with the unitary and causal approach based on the chiral Lagrangian
of Refs.~\cite{Gasparyan:2011yw,Gasparyan:2010xz}.
Although the two approaches are quite different in construction, they agree well
below  the pion-production threshold and start deviating slightly when the energy increases.
The small difference between the two calculations at higher energies can be used to estimate
the theoretical uncertainty in calculations of Compton scattering observables.
\newline
\noindent
A second cross check was  performed in Ref.~\cite{Pasquini:2010zr}, by comparing the ``experimental''
spin-dependent forward Compton scattering amplitude
(as constructed from the available data on the helicity-dependent absorption cross sections )
with the DR calculation obtained from different phenomenological inputs
for the pion-photoproduction multipoles.
The  spin-dependent  forward Compton scattering amplitude can be parametrized as
\begin{equation}
g(\nu)=-\frac{e^2\kappa_N^2}{8\pi M_N^2}\nu+\gamma_0^{\rm{dyn}}(\nu)\nu^3 \, ,
\end{equation}
where the leading term is due to intermediate nucleon states (Born terms) and the higher-order contribution
is described  by the energy-dependent  (dynamic) forward spin polarizability (FSP) $\gamma_0^{\rm{dyn}}(\nu)$.
This polarizability is a complex quantity which obeys
 the following DRs:
\begin{eqnarray}
{\rm {Re}}[\gamma_0^{\rm{dyn}}(\nu)] &=& \frac{1}{4\pi^2}\,{\mathcal{P}}\,\int_{\nu_0}^{\infty}\,\frac{\sigma_{1/2}(\nu')-\sigma_{3/2}(\nu')}
{\nu' ({\nu'}^2-\nu^2)}\,d\nu' \, ,\label{eq:dyn-FSP-real}
\\
{\rm {Im}}[\gamma_0^{\rm{dyn}}(\nu)] &=& \frac{\sigma_{1/2}(\nu)-\sigma_{3/2}(\nu)}{8 \pi \nu^2} \, .\label{eq:dyn-FSP-im}
\end{eqnarray}
For $\nu<\nu_0$ the imaginary part vanishes, and the dynamical FSP
has a LEX with the leading-order contribution given by the forward polarizability $\gamma_0$.
The ``experimental'' dynamical FSP is compared in Fig.~\ref{fig3} with the predictions based on
different phenomenological parametrizations of the pion photoproduction multipoles~\cite{Hanstein:1997tp,Arndt:2002xv,Drechsel:2007if,DMT:2001}.
The onset of $S$-wave pion production
at $E_{\gamma}=\nu_0$ leads to a strong cusp effect in the real part. The rapid increase of the dynamic FSP
from its static value $-0.90\cdot 10^{-4}$ ${\rm {fm}}^4$ at $E_{\gamma}=0$ to about $6\cdot 10^{-4}$ ${\rm {fm}}^4$
at pion threshold clearly shows the necessity to analyze Compton scattering within the framework of
dispersion analysis. In particular, such an approach is prerequisite in order to determine the 4 spin
polarizabilities, which yield significant contributions to the cross section only for photon energies above 80~MeV where the LEX
does not apply.
Except for the minimum of ${\rm {Re}}[\gamma_0^{\rm{dyn}}]$ near $E_{\gamma}=0.23$~GeV, the experimental analysis is in
very good agreement with the predictions of the presented models. The deviation in the minimum is not too surprising,
because this comes about by a delicate balance between the positive contribution
from the $S$-wave background
and the negative contribution from the low-energy tail of the $\Delta (1232)$ resonance.
\section{VCS and Generalized Polarizabilities}
\label{section:4}
The VCS process on the proton is accessed through
the $e p \to e p \gamma$ reaction.
In this process, the final photon can be emitted
either by the proton, which is referred to as the fully virtual
Compton scattering (FVCS) process, or by
the lepton, which is referred to as the Bethe-Heitler (BH) process.
The amplitude $T^{ee'\gamma}$ of the $e p \to e p \gamma$
reaction is given by the coherent sum of the BH and the FVCS process~:
$
T^{ee'\gamma}=T^{\rm{BH}}+T^{\rm{FVCS}}.
$
The BH amplitude $T^{\rm{BH}}$ is exactly calculable from QED if one knows
the nucleon e.m. form factors. The FVCS amplitude
$T^{\rm{FVCS}}$ contains, in the one-photon exchange approximation, the VCS
subprocess $\gamma^* p \to \gamma p$.
The VCS amplitude can be further decomposed into a Born
and a non-Born contribution. In the Born process, the virtual photon is
absorbed on a nucleon and the intermediate state remains a nucleon,
whereas the non-Born process contains all nucleon excitations
and meson-loop contributions.
The separation between Born and non-Born parts is performed as described in
Ref.~\cite{Guichon:1995pu,Scherer:1996ux}, to which we refer for the details.
The behavior of the non-Born VCS contribution
at low outgoing-photon energy $E'_\gamma$
but at arbitrary three-momentum
of the virtual photon, can be parametrized by 6 generalized
polarizabilities (GPs), which depend on the virtuality $Q^2$ transferred by the virtual photon.
The GPs are denoted by $P^{(\mathcal{M}' \, L',\, \mathcal{M} \,L)S}(Q^2)$
\cite{Guichon:1995pu,Drechsel:1997xv,Drechsel:1996ag}.
In this notation, $\mathcal{M}$ ($\mathcal{M}'$) refers to the
electric $(E)$, magnetic $(M)$ or longitudinal $(L)$ nature of the initial
(final) photon, $L$ ($L'$) represents the angular momentum of the
initial (final) photon, and $S$ differentiates between the
spin-flip ($S=1$) and non spin-flip ($S=0$)
character of the transition at the nucleon side.
Assuming that the emitted real photons have low energies, we may use
the dipole approximation ($L' = 1$). For a dipole transition in the
final state, angular momentum and parity conservation lead to 10 GPs
\cite{Guichon:1995pu}. Furthermore, it has been shown~\cite{Drechsel:1997xv} that nucleon crossing
combined with charge conjugation symmetry of the VCS amplitudes
provides 4 additional constraints among the 10 GPS, which leaves only 6 independent GPs.
\newline
\indent
In the limit of $E'_\gamma\rightarrow 0$, the VCS experiments
can be analyzed in terms of LEXs, as proposed in Ref.~\cite{Guichon:1995pu}.
These expansions are based on a low-energy theorem stating that the radiative
amplitude for point-like particles diverges like $1/E'_\gamma$ for $E'_\gamma \rightarrow 0$,
whereas the dispersive amplitude vanishes like $E'_\gamma$ in that limit.
The interference between the  $1/E'_\gamma$ contribution of BH plus Born amplitudes
and the leading term of the non-Born amplitude, can be expressed by 3
structure functions depending on the GPs,
\begin{eqnarray}
P_{LL} &=& - 2\sqrt{6} \, M_N  G_E  P^{( {L1,L1} )0} \,,
\nonumber \\
P_{TT}&=& - 3  G_M \frac{q^2}{\tilde {q}_0} \left( P^{(M1,M1)1}-\sqrt{2}\,
\tilde {q}_0  P^{(L1,M2)1} \right) \,,
\label{eq:4.2.19} \\
P_{LT}&=& \sqrt{\frac{3}{2}}\, \frac{M_N q}{Q} G_E  P^{(M1,M1)0}+\frac{3}{2}
\frac{Q q}{\tilde {q}_0}G_M P^{(L1,L1)1}\,, \nonumber
\end{eqnarray}
with $G_E$ and $G_M$ the electric and magnetic nucleon
form factors.\\
However, since the sensitivity of the VCS cross sections to the GPs
grows with the photon energy, it is
advantageous to go to higher photon energies, provided one can keep the
theoretical uncertainties under control when approaching and crossing the pion
threshold. The situation can be compared to RCS as described in
Sec. \ref{section:2}, where it was shown that a DR formalism is prerequisite
to extract the polarizabilities in the energy region
above pion threshold where the observables are generally more sensitive to the GPs.
In order to set up DRs for VCS, we describe the VCS tensor in terms of 12 independent amplitudes $F_i$ ($i=1,\dots ,12$)
which are free of kinematical singularities and constraints and even under crossing.
These amplitudes depend on the 3 variables $\nu, \, t $ and $Q^2$, with
$\nu =  E_{\gamma} + (t - Q^2)/(4 M_N) \,,$ and
$t =2 E'_{\gamma} \, (\cos\,\theta_{\rm{lab}}\,\sqrt {E_{\gamma}^2 + Q^2} -E_{\gamma}) - Q^2 $.
\begin{figure}
\vspace{-1cm}
\begin{center}
\includegraphics[width = 8.cm]{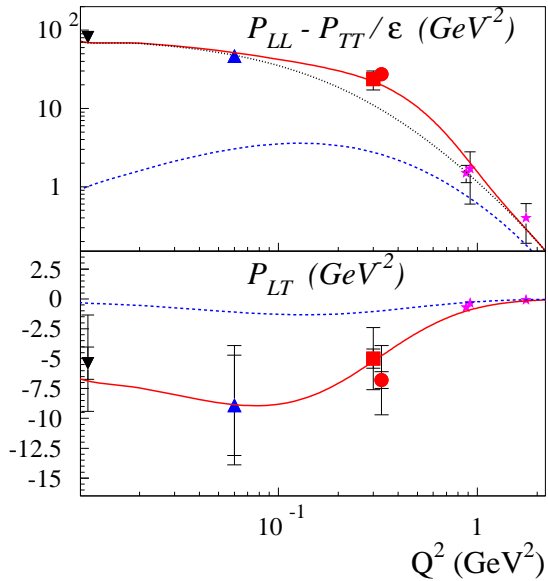}
\end{center}
\vspace{-0.5cm}
\caption{The structure functions describing unpolarized VCS on a proton compared with the data
from  MAMI (circles~\cite{Roche:2000ng}, squares~\cite{Janssens:2008qe}),
MIT-Bates (up triangles~\cite{Bourgeois:2006js}), and JLab (stars~\cite{Laveissiere:2004nf}).
The RCS data~\cite{OlmosdeLeon:2001zn} are shown by the (black) down triangles (slightly displaced in $Q^2$).
All the lines are shown for $\varepsilon = 0.645$ and based on the parameterizations of
Eqs.~(\ref{eq:gpmagn}) and (\ref{eq:gpelec}) for the proton GPs.
The lines in the upper panel correspond to different sets of parameters,
dotted (black): $\Lambda_\alpha = 0.7$~GeV,  $C_\alpha = 0$ (GP I),
solid (red): $\Lambda_\alpha = 0.7$~GeV,  $C_\alpha = -150$~GeV$^{-7}$ (GP II).
The solid (red) line in the lower panel is obtained for $\Lambda_\beta = 0.5$~GeV.
The dashed (blue) lines in both panels show the contributions of the spin GPs.\label{fig4}}
\end{figure}
Assuming further an appropriate analytic
and high-energy behavior, these amplitudes fulfill unsubtracted DRs
in the variable $\nu$ and at fixed $t$ and $Q^2$,
\begin{eqnarray}
\mathrm{Re} \,F_i^{\mathrm{nB}}(Q^2, \nu, t) = F_i^{\mathrm{pole}}(Q^2,
\nu, t)- F_i^{\mathrm{B}}(Q^2, \nu, t)\nonumber \\
+\frac {2} {\pi} \, \mathcal {P} \int_{\nu_{0}}^{\infty} d\nu' \, \frac{\nu' \,
\mathrm{Im} F_i(Q^2, \nu',t)} {\nu'^2 - \nu^2}\, , \label{eq:4.2.11}
\end{eqnarray}
where  the pole amplitudes
$F_i^{\mathrm{pole}}$ are obtained from the Born amplitudes at the pole position,
that is, with all numerators evaluated at the pole. Furthermore, Im~$F_i$ are
the discontinuities across the $s$-channel cuts, starting at the
pion-production threshold, which can be calculated from the absorption
cross sections due to $\pi N$, $\pi \pi N$, and heavier hadronic states. As long
as we are interested in the energy region up to the $\Delta(1232)$, we may
saturate the $s$-channel dispersion integral by the $\pi N$ contribution,
choosing $\nu_{\rm{max}} \approx 1.5$~GeV as upper limit of integration and using
empirical information from pion photo- and electroproduction as encoded in the
MAID2007 parametrization~\cite{Drechsel:2007if}.
However, Eq.~(\ref{eq:4.2.11}) is only valid if the amplitudes drop
fast enough such that the integrals converge. The high-energy behavior of the
amplitudes $F_i$ was investigated by \cite{Pasquini:2001yy} in the Regge
limit ($\nu \rightarrow \infty$, $t$ and $Q^2$ fixed), where it was found
that the dispersion integrals diverge for two
amplitudes, $F_1$ and $F_5$.
The asymptotic contributions to these  amplitudes are described by
the $t$-channel exchange of $\sigma$ and $\pi^0$ mesons, respectively,
and other effects such as many-pion and heavier intermediate states in the $s$-channel integral.
All these effects beyond the dispersive $\pi N$ contribution are parametrized in terms of
asymptotic contributions. In particular, the asymptotic part of the magnetic GP is described
by a dipole function
\begin{eqnarray}
P^{(M1,M1)0}_{asy}(Q^2) &=& P^{(M1,M1)0}_{asy}(0) / (1 + Q^2 / \Lambda_\beta^2)^2 .  \;\;\;\;\;
\label{eq:gpmagn}
\end{eqnarray}
To describe the electric GP, we allow for an asymptotic part consisting of
a sum of a dipole and a Gaussian, in the same vein as the parametrization
proposed in \cite{Friedrich:2003iz} for the nucleon form factors,
\begin{eqnarray}
P^{(L1,L1)0}_{asy}(Q^2) &=& P^{(L1,L1)0}_{asy}(0) / (1 + Q^2 / \Lambda_\alpha^2)^2
+ C_\alpha \, Q^4 \, e^{- (Q^2 - 0.15) / 0.15}.
\label{eq:gpelec}
\end{eqnarray}
The values at the real-photon point are fixed by the difference between
the dispersive $\pi N$ contribution and the empirical information
for the proton electric and magnetic polarizabilities as obtained from RCS
experiments~\cite{OlmosdeLeon:2001zn}, which leads to
$P^{(L1,L1)0}_{asy}(0) = -14.37$~GeV$^{-3}$, and $P^{(M1,M1)0}_{asy}(0) = 21.82$~GeV$^{-3}$.
The remaining three parameters $\Lambda_\beta, \Lambda_\alpha$, and $C_\alpha$ describe
the $Q^2$ dependence of the asymptotic parts of the spin-independent GPs and can be determined
by a fit to the available VCS data.
In the upper panel of Fig.~\ref{fig4}, we show the comparison with the experimentally measured unpolarized
structure functions $P_{LL} - P_{TT} / \varepsilon$ ($P_{LT}$), which is proportional
to the electric (magnetic) GPs respectively, up to a small spin-GP contribution (dashed lines).
The magnetic GP is essentially given by $P_{LT}$ shown in the lower panel of {Fig.~\ref{fig4}.
The figure shows that all the data can be well described by $\Lambda_\beta = 0.5$~GeV.
For the electric GP, a fit to the MIT-Bates and JLab data is obtained for $C_\alpha = 0$,
and $\Lambda_\alpha = 0.7$~GeV (denoted by parameterization GP I).
However, this does not describe the MAMI data at intermediate $Q^2$, which require an additional structure, parameterized through
the Gaussian term in Eq.~(\ref{eq:gpelec}). A good description of all available data is obtained for
$\Lambda_\alpha = 0.7$~GeV, and $C_\alpha = -150 \;  \mathrm{GeV}^{-7}$
(denoted by parameterization GP II).
\newline
\noindent
\begin{figure}
\vspace{1 truecm}
\begin{center}
\includegraphics[width = 0.65\columnwidth]{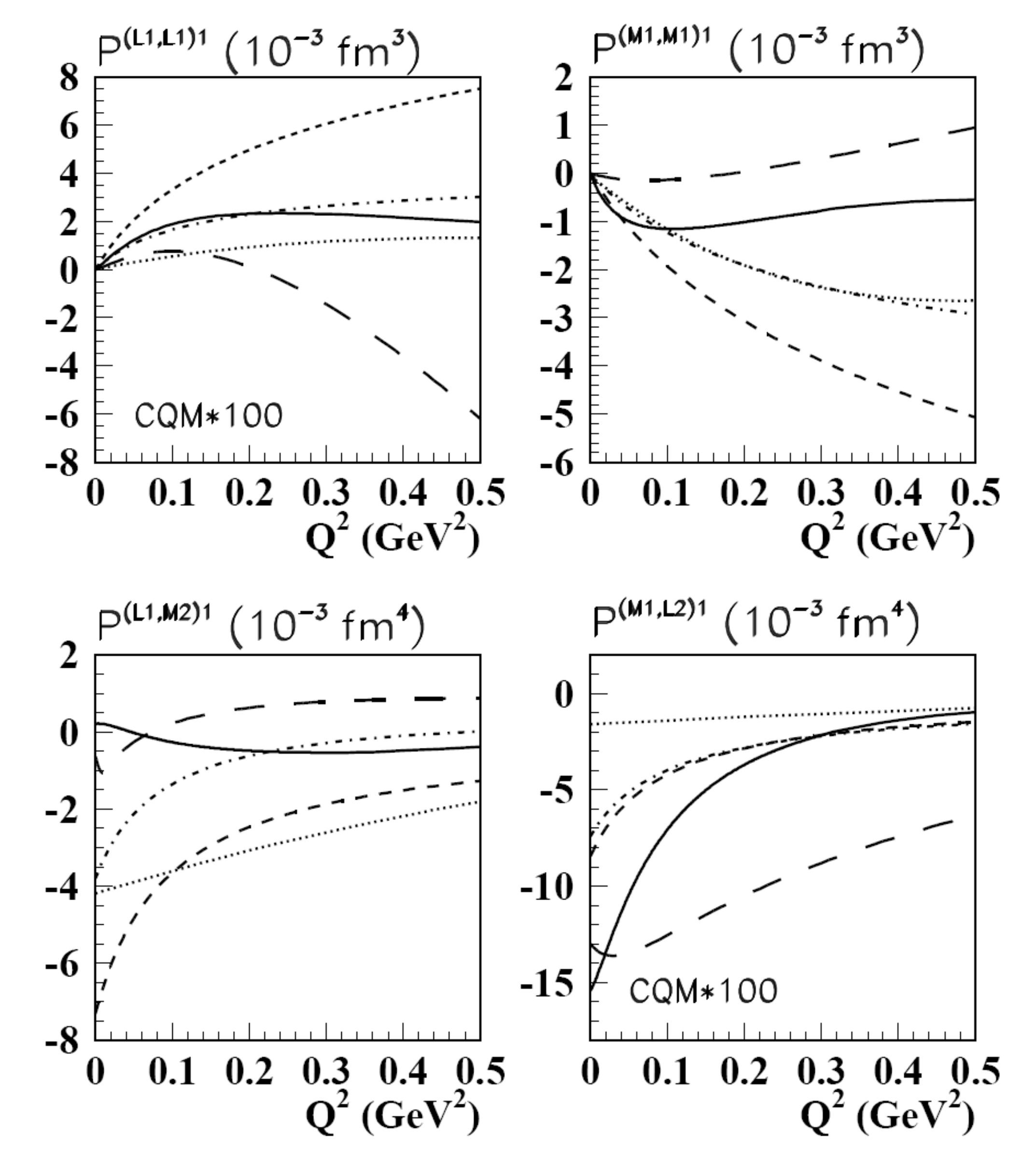}
\end{center}
\caption{The spin-flip GPs without the $\pi^0$-pole contribution.
The lines shown correspond to the predictions from several
calculations. Solid: dispersive $\pi N$ contribution
\cite{Pasquini:2001yy}, short-dashed: ${\mathcal O}(p^3)$ HBChPT
\cite{Hemmert:1996gr,Hemmert:1999pz}, long-dashed: ${\mathcal O}(p^4)$ HBChPT
\cite{Kao:2002cn,Kao:2004us}, dashed-dotted: linear $\sigma$ model
\cite{Metz:1996fn}, dotted: non-relativistic CQM \cite{Pasquini:2000ue}.
For visibility, the tiny CQM results for $P^{(L 1,L 1)1}$ and $P^{(M 1,L 2)1}$
are multiplied by a factor 100.\label{fig5}}
\vspace{-0.5 truecm}
\end{figure}
The GPs  were also calculated in HBChPT at $\mathcal{O}(p^3)$~\cite{Hemmert:1996gr,Hemmert:1999pz} and at $\mathcal{O}(p^4)$~\cite{Kao:2002cn,Kao:2004us}.
Although the HBChPT and DR results agree qualitatively in the description of the
unpolarized VCS structure functions, there are large differences in the spin sector.
Figure~\ref{fig5} shows the predictions for the spin GPs for several different approaches. The comparison
demonstrates that a satisfying theoretical description of the spin-flip GPs over a large range
in $Q^2$ is still a challenging task. This calls for VCS experiments which are sensitive to the spin GPs.
Two types of such experiments can be envisaged: ($i$) unpolarized VCS with variation of the transverse photon
polarization $\epsilon$ in order to separate the response functions $P_{LL}$
and $P_{TT}$ and ($ii$) double-polarization experiments to access new structure
functions directly related to the spin-flip GPs~\cite{Vanderhaeghen:1997bx}.
A first test experiment for double-polarization observables
was performed at MAMI~\cite{MAMI:VCS-DP}.
Furthermore, these structure coefficients are also prominent input in calculating hadronic corrections to the Lamb shift~\cite{Carlson:2011zd} and the hyperfine splitting in hydrogen~\cite{Nazaryan:2005zc}.
\section{Spatial representation of GPs}
\label{section:5}
In the last years, generalized parton distributions~\cite{Burkardt:2000za,Diehl:2005jf}
and form factors~\cite{Miller:2007uy,Carlson:2007xd} have
been discussed as a tool to access the distributions of partons in the transverse plane,
and first calculations of these spatial distributions have been performed in lattice QCD~\cite{:2003is}
and hadronic models~\cite{Pasquini:2007xz}.
Most recently, it has also been shown how the concept of GPs can be used to describe
the spatial deformation of the charge and magnetization densities if the nucleon is exposed
to an external quasi-static electric field~\cite{L'vov:2001fz,Gorchtein:2009qq}.
\newline
\noindent
In order to arrive at a spatial representation of the information contained in the GPs,
we consider the VCS process in a  light-front frame~\cite{Gorchtein:2009qq}.
The VCS tensor is defined as
\begin{eqnarray}
H^{\mu \nu} = -i \int d^4 x \, e^{- i q \cdot x} \langle p^\prime, \lambda^\prime_N |
T \left[ J^\mu(x), J^\nu(0) \right] | p,   \lambda_N \rangle,
\label{eq:tordered}
\end{eqnarray}
with $\lambda_N$ and $p$ ($\lambda^\prime_N$ and $p'$) the helicity and four-momentum of the initial (final) nucleon.
 We indicate the (large) light-front + component by $P^+=(p^++p'^+)/2$
(defining $a^\pm \equiv a^0 \pm a^3$), and choose the symmetric frame by requiring that
$\Delta = p^\prime - p$ is purely transversal, i.e. $\Delta^+ = 0$.
In the light-front frame, the + component of the current $J^\mu$ in~(\ref{eq:tordered})
is a positive-definite operator for each quark flavor,
allowing for a light-front charge density interpretation.
In the low-energy limit $\nu\rightarrow 0$, the VCS light-front helicity amplitude $H(\lambda^\prime_\gamma, \lambda^\prime_N , \lambda_N)
\equiv  \varepsilon^{\, \prime *}_\nu (\lambda^\prime_\gamma) \, H^{+ \nu}$
can be described in terms of GPs.
In particular, we  consider the transverse polarization component of the
outgoing photon corresponding to an electric field $
\vec E \sim i \, E^{\prime}_\gamma  \, \vec \epsilon^{\, \prime}_\perp $.
Any system of charges will respond to such an applied electric field through
an induced polarization $\vec P_0$, which will be forced to align with the external field
such as to minimize the energy $ - \vec E \cdot \vec P_0$.
The linear response in the outgoing-photon energy  of the helicity averaged VCS amplitude therefore
allows one to define an induced polarization $\vec P_0$ as
\begin{equation}
\, \vec \varepsilon_\perp^{\, \prime *} (\lambda^\prime_\gamma) \cdot \vec P_0
\equiv \frac{(1 + \tau)}{(2 P^+)}
\frac{\partial}{\partial \nu}
H \left(\lambda^\prime_\gamma, \lambda_N, \lambda_N \right) \big|_{\nu = 0} \, ,
\label{eq:indpol}
\end{equation}
where $\tau \equiv Q^2 / (4 M_N^2)$.
\begin{figure}[t]
\begin{center}
\vspace{-1. cm}
\includegraphics[width = 6.1cm]{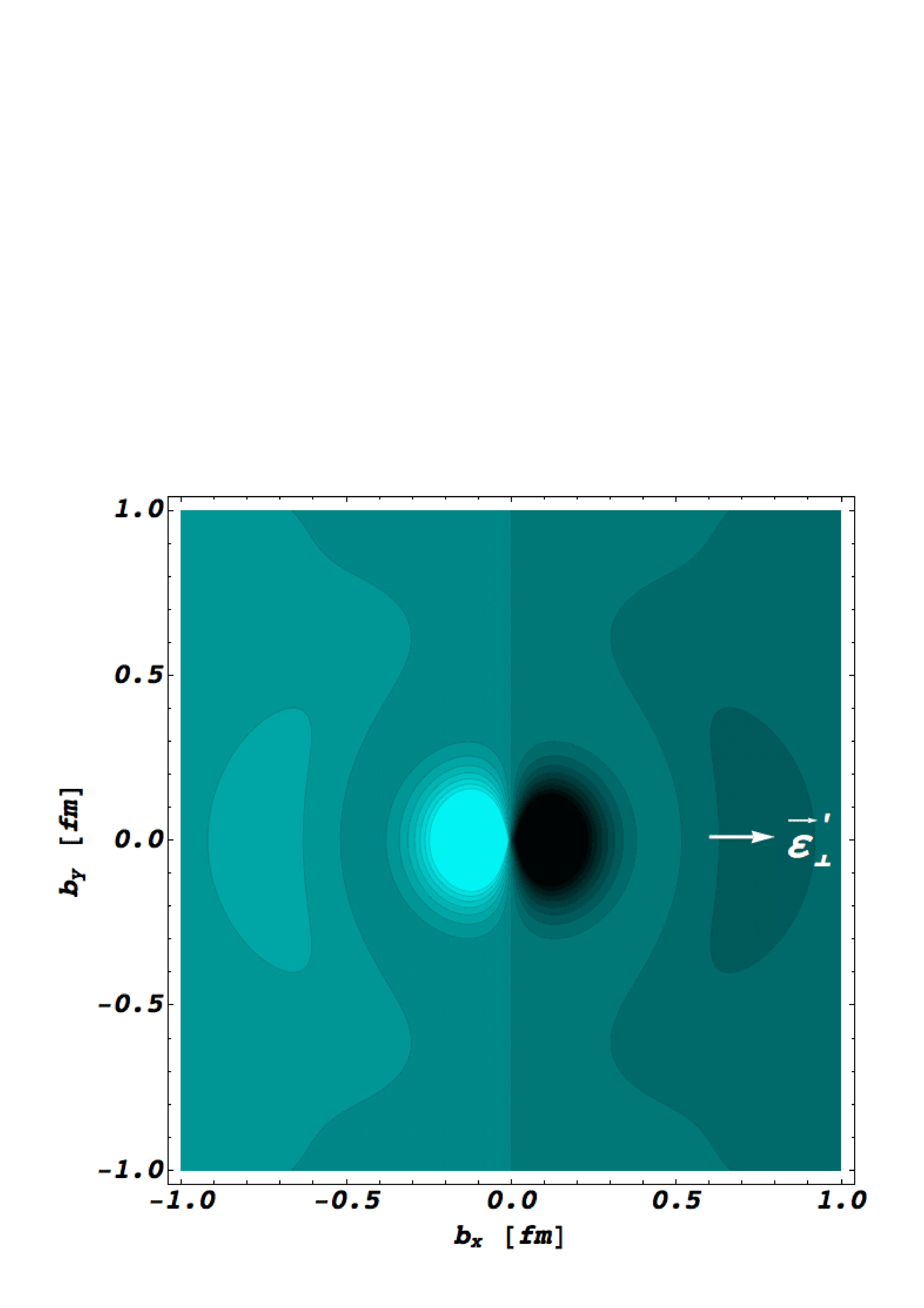}
\includegraphics[width = 6.1cm]{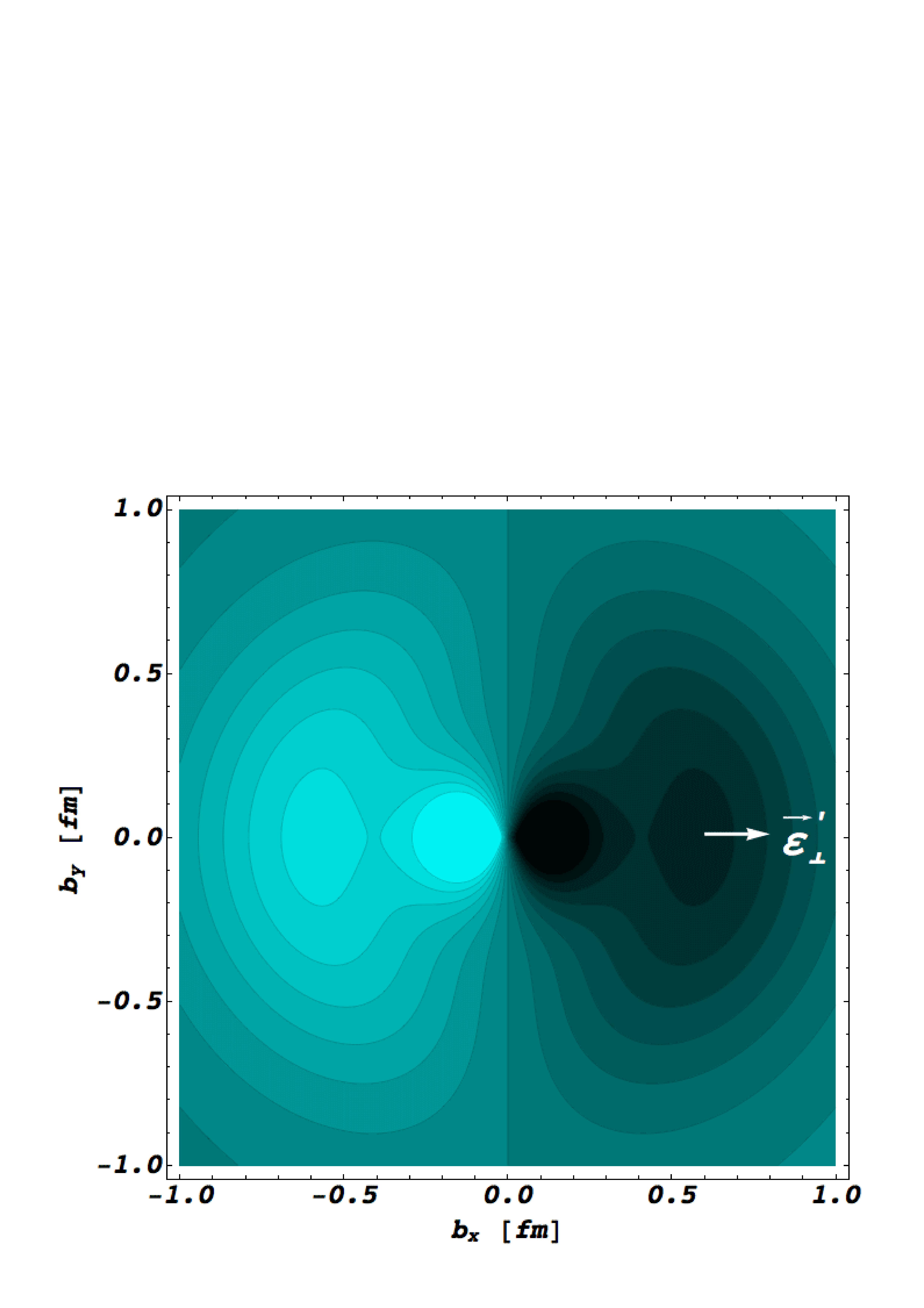}
\end{center}
\caption{The induced polarization $P_0^x$ in a proton of
definite light-cone helicity if exposed to an e.m. field with photon polarization along the $x$-axis,
as indicated. The left (right) panel is obtained for GP~I (GP II), see caption of Fig.~\ref{fig4}.
The light (dark) regions correspond to the largest (smallest) values. \label{fig6}}
\end{figure}
\begin{figure}[t]
\begin{center}
\vspace{-1 truecm}
\includegraphics[width = 6.1cm]{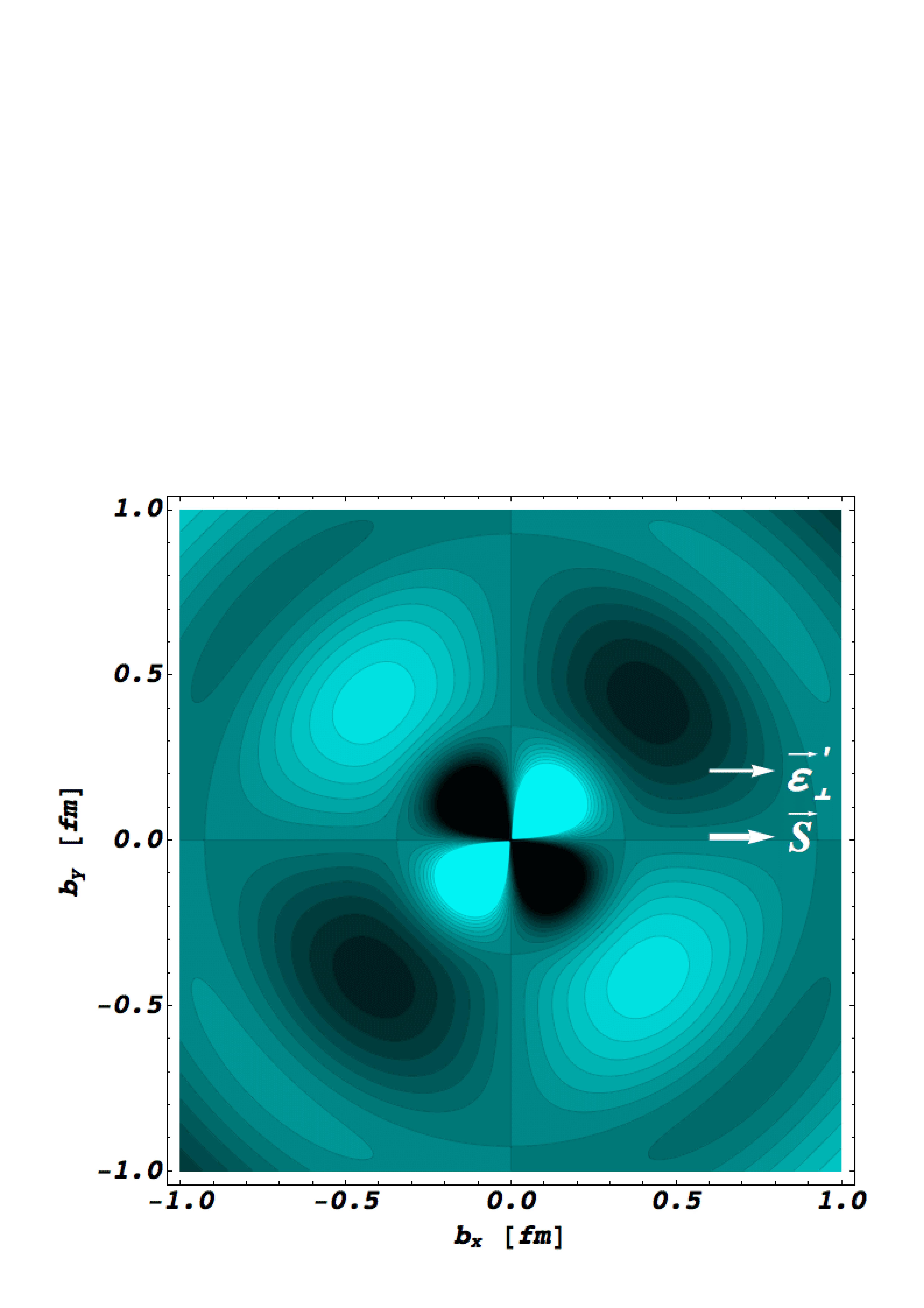}
\includegraphics[width = 6.1cm]{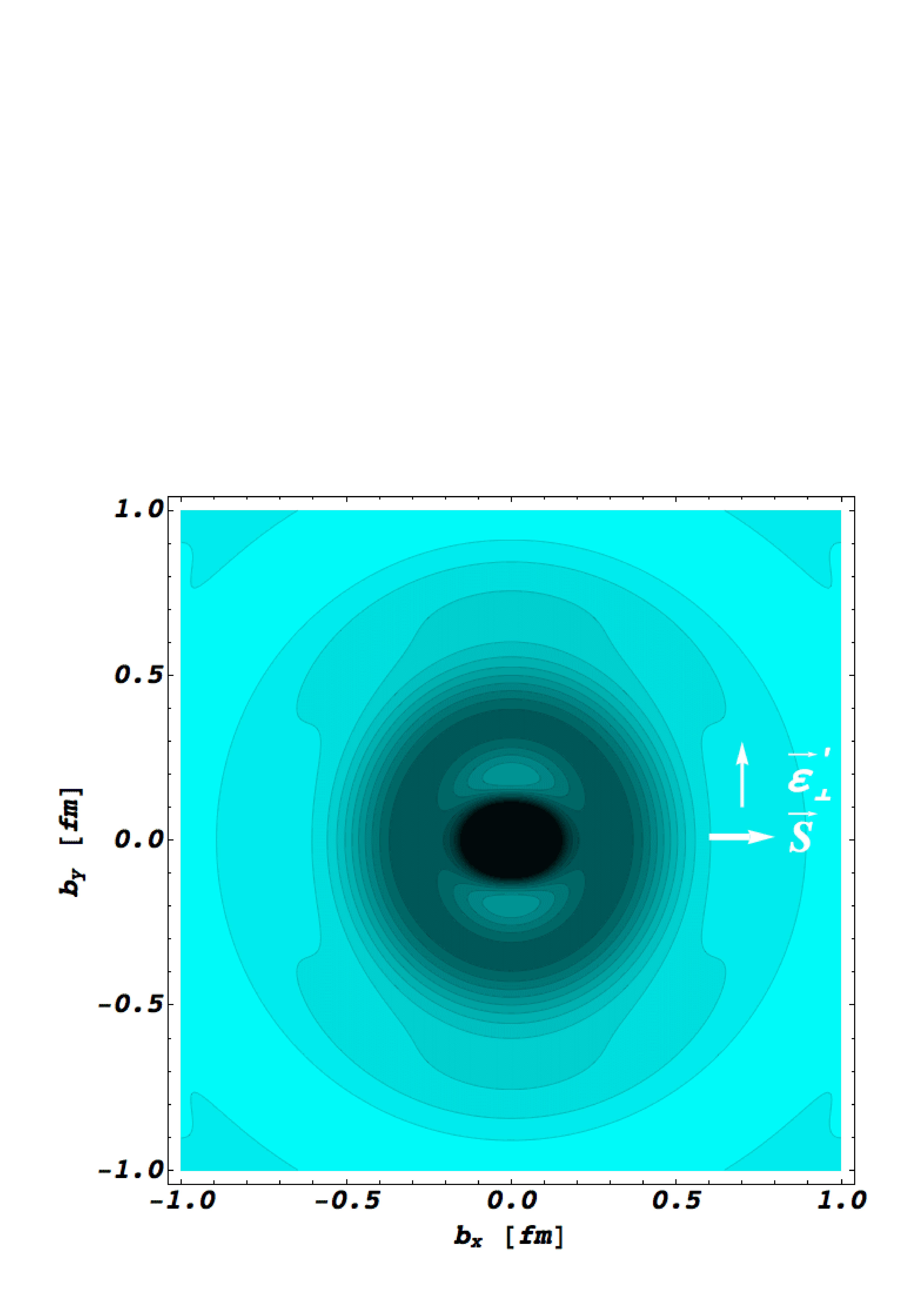}
\end{center}
\caption{ The nduced polarization density in a proton, with spin $\vec S$ oriented along the $x$-axis, when submitted to an e.m. field.
The left (right) panels is  for $P_T^x - P_0^x$ ($P_T^y - P_0^y$),
and correspond with photon polarization along the $x$-axis ($y$-axis) as indicated.
The light (dark) regions correspond to the
largest (smallest) values using parameterization GP II, see caption of Fig.~\ref{fig4}.\label{fig7}}
\end{figure}
By Fourier transforming  $\vec P_0$, we obtain the spatial distortion
of the charge density in an unpolarized nucleon as induced by the external electric field,
\begin{equation}
\vec P_0(\vec b) = \hat b \, \int_0^\infty \frac{d Q}{(2 \pi)} \, Q  \, J_1(b \, Q) \,  A(Q^2),
\label{eq:indpolnoflipb}
\end{equation}
where $\vec b$ is the transverse position, $b = | \vec b|$, and $\hat b = \vec b / b$ and $A(Q^2)$
can be expressed in terms of the scalar and spin GPs.
The dipole pattern described by Eq.~(\ref{eq:indpolnoflipb}) is shown in Fig.~\ref{fig6}
for the parameterizations GP I and GP II in the left and right panels, respectively. The comparison shows
that the enhancement at intermediate $Q^2$ in GP II (second term in Eq.~(\ref{eq:gpelec})) gives rise to a much larger transverse extension
of the induced polarization cloud. Forthcoming VCS experiments planned at MAMI~\cite{MAMI:VCS} are conceived
to study the $Q^2$ dependence of the GPs in more detail. In this way, the experiment is expected to verify or disprove
the large distance structure predicted by model GP II.
\newline
\indent
Analogously, we can define the linear response to an external
quasi-static e.m. dipole field if the nucleon is in an eigenstate of transverse spin,
$\vec S_\perp \equiv \cos \phi_S \hat e_x + \sin \phi_S \hat e_y$,
with $\phi_S$ the angle indicating the spin vector direction. The dependence of the
induced polarization $\vec P_T$ on the transverse position is given by
\begin{eqnarray}
\vec P_T (\vec b) &=&  \vec P_0(\vec b)
- \hat b \, \left( \vec S_\perp \times \vec e_z \right) \cdot \hat b \,
\int_0^\infty \frac{d Q}{(2 \pi)} \, Q  \, J_2(b Q) \,  B \nonumber \\
&+&  \left( \vec S_\perp \times \vec e_z \right)
\int_0^\infty \frac{d Q}{(2 \pi)} Q  \left[ J_0(b Q)   C + \frac{J_1(b Q)}{bQ}  B
\right],
\end{eqnarray}
displaying dipole, quadrupole and monopole patterns with weights
$B$, and $C$  given by combinations of scalar and spin GPs.
\newline
\indent
In Fig.~\ref{fig7} we show the spatial distributions in the induced polarization
for a proton of transverse spin (chosen along the $x$-axis) for parameterization GP II.
The component  $P_T^x - P_0^x$ displays a quadrupole pattern with pronounced strength around $0.5$~fm
due to the electric GP, whereas the component $P_T^y - P_0^y$ shows an additional monopole pattern,
which is dominated by the $\pi^0$-pole contribution.
\section{Conclusions}
\label{section:6}

In this contribution we have discussed real and virtual Compton scattering off the proton as
powerful tools to extract information on the proton polarizabilities.
In the case of RCS, the static scalar polarizabilities are now known with relatively small error bars,
whereas our knowledge on the spin-dependent sector is as yet incomplete.
Only two combinations of the spin polarizabilities have been determined as yet. However, the
forthcoming double-polarization experiments at MAMI will allow us to disentangle all 6 spin-polarizabilities
individually. The spin polarizabilities of the nucleon are fundamental
structure constants of the nucleon, just as shape and size of this
strongly interacting many-body system. Therefore, the ongoing experimental programs are both timely and important
for our understanding of the nucleon structure. These activities will provide stringent precision tests
for both the existing predictions of effective field theories and future results expected
from the lattice-gauge community for the polarizability of the nucleon.
\newline
\indent
Moreover, the generalized polarizabilities from VCS processes allow us to
map out the spatial dependence of the induced polarizations in an external e.m. field.
The existing field theoretical formalism to extract
light-front densities from nucleon form factor data has been recently extended to the deformations of quark charge densities
under the influence of an applied e.m. field. It has been shown that the available proton electric GP data
yield a pronounced structure in the induced polarization density at large transverse distances of $0.5 - 1$~fm.
At $Q^2$ values smaller than 0.1 GeV$^2$, chiral effective field theory was found to describe the VCS data quite well,
which highlights the role of pions in the nucleon structure. Such description can however not be applied
at intermediate and large $Q^2$ values. This transition region is dominated by nucleon resonance structure,
which can be described by dispersion relations. Forthcoming VCS precision experiments at MAMI in this intermediate $Q^2$
region will be capable to determine the spin polarizabilities in more detail and, in this way,
help to unravel the the distribution of quark charge and magnetization currents in the nucleon.
\begin{acknowledgement}
The authors are grateful to all the collaborators with whom many of the presented results were obtained.
We also acknowledge the support by the Collaborative Research Center 443
(SFB 443) of the Deutsche Forschungsgemeinschaft.
\end{acknowledgement}

\end{document}